\documentclass[%
superscriptaddress,
 amsmath,amssymb,
 aps,
]{revtex4-2}

\usepackage{graphicx}
\usepackage{dcolumn}
\usepackage{bm}
\usepackage{hyperref}

\begin{document}

\preprint{Physical Review C}

\title{\textbf{Elemental and angular fragmentation cross section measurements with the FOOT experiment} 
}%

\author{M. Dondi}
\email{Matilde.Dondi@bo.infn.it}
\affiliation{INFN Section of Bologna, Bologna, Italy}
\affiliation{University of Bologna, Department of Physics and Astronomy, Bologna, Italy}
\author{R. Ridolfi}%
\email{Riccardo.Ridolfi@bo.infn.it}
\affiliation{INFN Section of Bologna, Bologna, Italy}
\affiliation{University of Bologna, Department of Physics and Astronomy, Bologna, Italy}

\collaboration{on behalf of the FOOT Collaboration}


\begin{abstract}
The FOOT (FragmentatiOn Of Target) experiment was proposed to measure double differential nuclear fragmentation cross sections in angle and kinetic energy of the produced fragments in beam-target settings, interesting for hadrontherapy and space radioprotection applications. In particular, FOOT measures projectile and target fragmentations in the kinetic energy range between $200 \text{MeV/u}$ and $800 \text{MeV/u}$. In this contribution, differential cross section measurements of a $400 \text{MeV/u}$ $^{16}$O beam on a Carbon and a polyethylene target with data acquired at GSI (Darmstadt, Germany) beam accelerator facility are presented, along with the extraction of the first total fragmentation cross section for a Hydrogen target within the FOOT experiment.
\end{abstract}

\maketitle


\section{Introduction}
Hadrontherapy is an external radiation therapy used to treat cancer, employing charged particles. These release most of their energy just before stopping, in the Bragg peak region, the position of which depends on their initial energy. By tuning this parameter, it is possible to concentrate the energy deposition within the tumour region, minimizing damage to surrounding healthy tissues~\cite{ref:shulz}. At the energies of hadrontherapy, nuclear interactions between the beam and body nuclei can also occur, leading to the production of fragments. Their contribution to the biological dose distribution needs to be assessed~\cite{ref:durante}, but part of the needed cross sections have not been measured yet. Nuclear fragmentation measurements are also relevant in the field of space radioprotection, since space agencies have shown an increasing interest in human missions to deep space. Indeed, the main concern of this kind of missions is to protect astronauts from the harsh space radiation environment. To build accurate radiation models and provide effective shielding, data on nuclear interactions and possible nuclear fragmentation events are needed~\cite{ref:cucinotta}.
\section{The FOOT experiment}
\begin{figure}[h]
\centering
\includegraphics[width=0.7\textwidth]{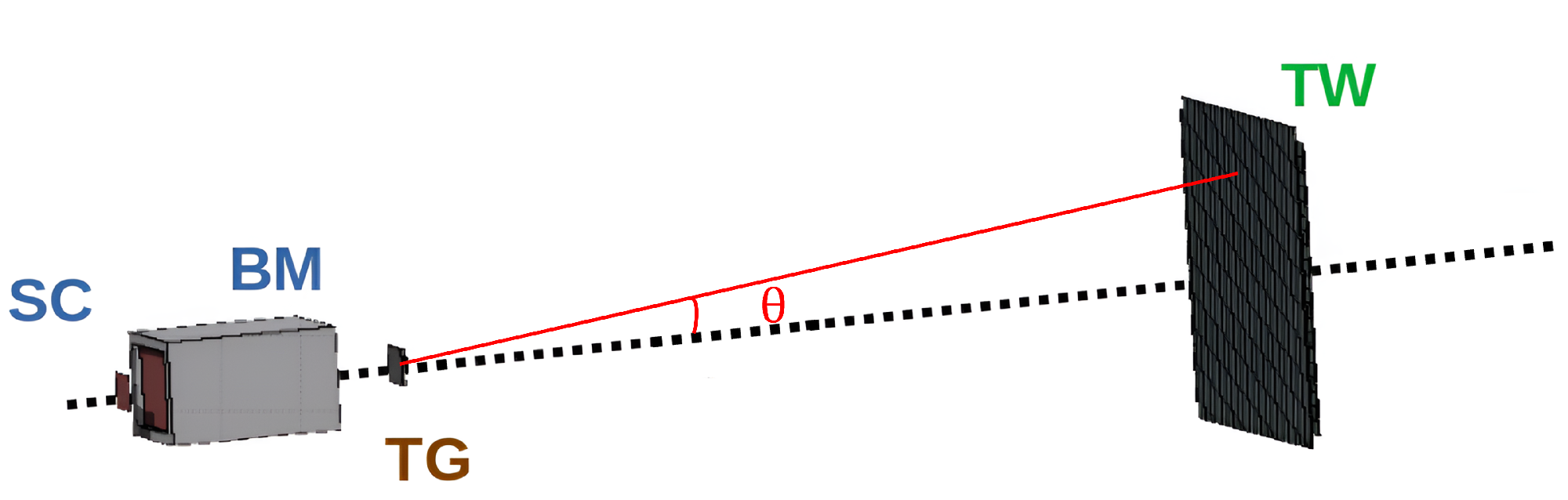}     
\caption{Schematic view of the GSI 2021 data taking setup.}
\label{f.setup}
\end{figure}

The FOOT (FragmentatiOn Of Target) experiment aims to provide precise measurements of double differential nuclear fragmentation cross section in angle and kinetic energy of the produced fragments, using different beams and targets, with an accuracy better than 5\%. In particular, FOOT measures projectile and target fragmentation in the energy range between $200\text{MeV/u}$ and $800\text{MeV/u}$ for hadrontherapy and space radioprotection applications. The experiment has two different setups: an emulsion setup optimized for light fragments with a wide angular distribution, and an electronic setup. In the GSI 2021 data taking, the FOOT electronic setup employed was not yet complete. It consisted of a first region in front of the target (TG) to track the beam with a $250\,\mu\text{m}$ thick foil of EJ-228 plastic scintillator and a drift chamber, referred to as the Start Counter (SC) and the Beam Monitor (BM) respectively, and almost two metres downstream of the target, the ToF-Wall (TW), two layers of twenty EJ-200 plastic scintillator bars, orthogonally oriented (fig.~\ref{f.setup}).
The SC measures the incoming ion flux, provides the trigger for the experiment, and marks the start time for the Time of Flight (ToF) and BM measurements. 
The BM reconstructs the beam trajectory before the target and the interaction point of the beam on the target. Information from this detector is also used to eliminate possible fragmentation events occurred in front of the target. 
The TW provides time information to retrieve the ToF and the energy loss of the fragments (\(\Delta E\)), from which their charge is reconstructed. More details about the FOOT setup and experimental program can be found in~\cite{ref:battistoni}.
\section{Analysis method}
This analysis aims to measure both total and angular fragmentation cross sections for various fragments, using data acquired at GSI with a $400\text{MeV/u}$ $^{16}$O beam on graphite (C) and polyethylene (C\(_2\)H\(_4\)) targets. Having these measurements in the same experimental conditions, it is also possible to extract the, otherwise hardly achievable, fragmentation cross sections on a Hydrogen target, using the subtraction method.\\
The charge Z of the produced fragments is identified using the ToF and \(\Delta E\) measured by the TW detector. These two quantities are related through a parametrization of the Bethe-Bloch formula as a function of the ToF. In a \(\Delta E\)-ToF plane, each fragment is plotted and an algorithm associates it with the Z corresponding to the closest Bethe-Bloch curve~\cite{ref:toppi}. \\
The main source of background, which was found to be out-of-target fragmentation events, cannot be actively removed without a tracking system, such as the one in the complete FOOT setup. For this reason, fragment yields measured on a run without target have been subtracted from fragment yields measured on runs with the target, normalizing the yields to the number of primaries. This strategy applied to data was firstly performed on MC samples generated with FLUKA validating the analysis procedure. Specifically, the cross section extracted using MC truth information was compared to the one reconstructed from the MC sample using the same analysis procedure applied to data. The MC was also used to extract purities and efficiencies.\\
To obtain a more precise angular reconstruction, dominated by the granularity of the TW bar crossing overlap of 2\(\times\)2\text{cm\(^2\)}, an angular unfolding procedure was implemented in the analysis. With this correction, the MC analysis showed a significant improvement in the cross section measurements, particularly for C (Z=6) and N (Z=7) fragments.
The formula for the angular cross sections, expressed as a function of the solid angle \(\Omega\), is:
\begin{equation}
\label{e.xs_angle}
\frac{\text{d} \sigma}{\text{d} \Omega}(Z,\theta) =\left( \frac{Y_\mathrm{TG}(Z,\theta)}{N_\mathrm{prim,TG}}-\frac{Y_{\mathrm{noTG}}(Z,\theta)}{N_{\mathrm{prim,noTG}}}\right)\frac{1}{\epsilon(Z,\theta)N_{\mathrm{TG}}\Delta\Omega}
\end{equation}
where \(\theta\) is the emission angle of the fragments, \(\frac{Y_\mathrm{TG}(Z,\theta)}{N_\mathrm{prim,TG}}\) and \(\frac{Y_\mathrm{noTG}(Z,\theta)}{N_\mathrm{prim,noTG}}\) are the fragment yields divided by the number of primary Oxygen ions for runs with and without the target respectively, \(N_{\mathrm{TG}}\) is the number of interaction centres per unit surface in the target, \(\Delta\Omega\) is the bin width and \(\epsilon\) is the efficiency. The efficiency is extracted from the MC sample and takes into account  the performance of the TW detector's clustering algorithm in reconstructing a fragment. 
The minimum angular bin width, chosen according to the TW granularity, is \(\simeq\) 0.6° and the binning for the different charges was chosen on the basis of the available statistics at higher angles for heavier fragments, which are characterized by a narrow distribution.
The total cross section can be extracted from the angular one integrating within the 5.7° angular acceptance of the TW detector. \\
Starting from the cross section results obtained for polyethylene and Carbon targets, it was also possible to extract the cross section for a Hydrogen target as:
\begin{equation}
 \label{e.xs_p}
\sigma[\mathrm{H}] = \frac{1}{4}(\sigma[\mathrm{C}_2\mathrm{H}_4]-2\sigma[\mathrm{C}]) 
\end{equation}
where \(\sigma[\mathrm{C}_2\mathrm{H}_4]\) and \(\sigma[\mathrm{C}]\) are the cross sections for polyethylene and Carbon targets respectively. The same formula holds for the angular cross sections.
\section{Results}
\begin{figure}[h]
\centering
\begin{minipage}{0.49\textwidth}
\centering
\includegraphics[width=\textwidth]{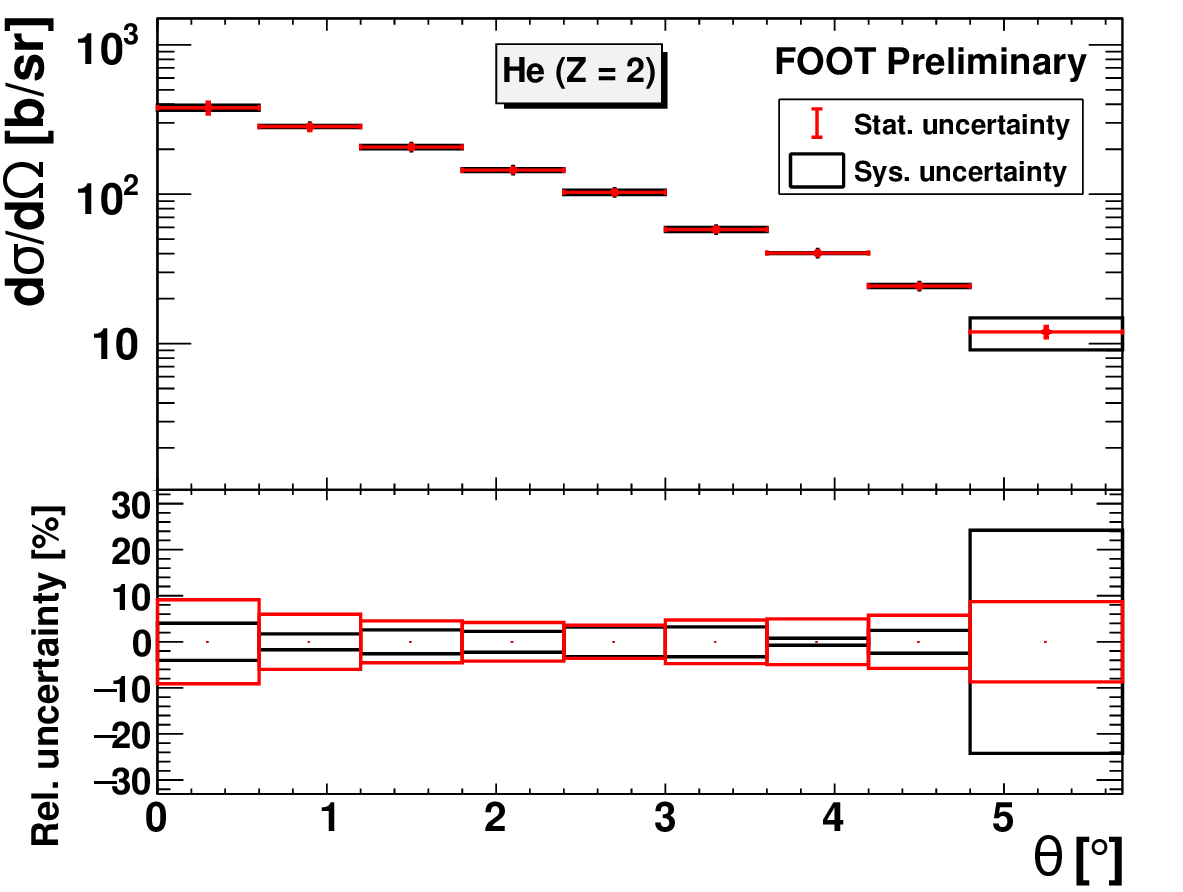}
\end{minipage}
\hfill
\begin{minipage}{0.49\textwidth}
\centering
\includegraphics[width=\textwidth]{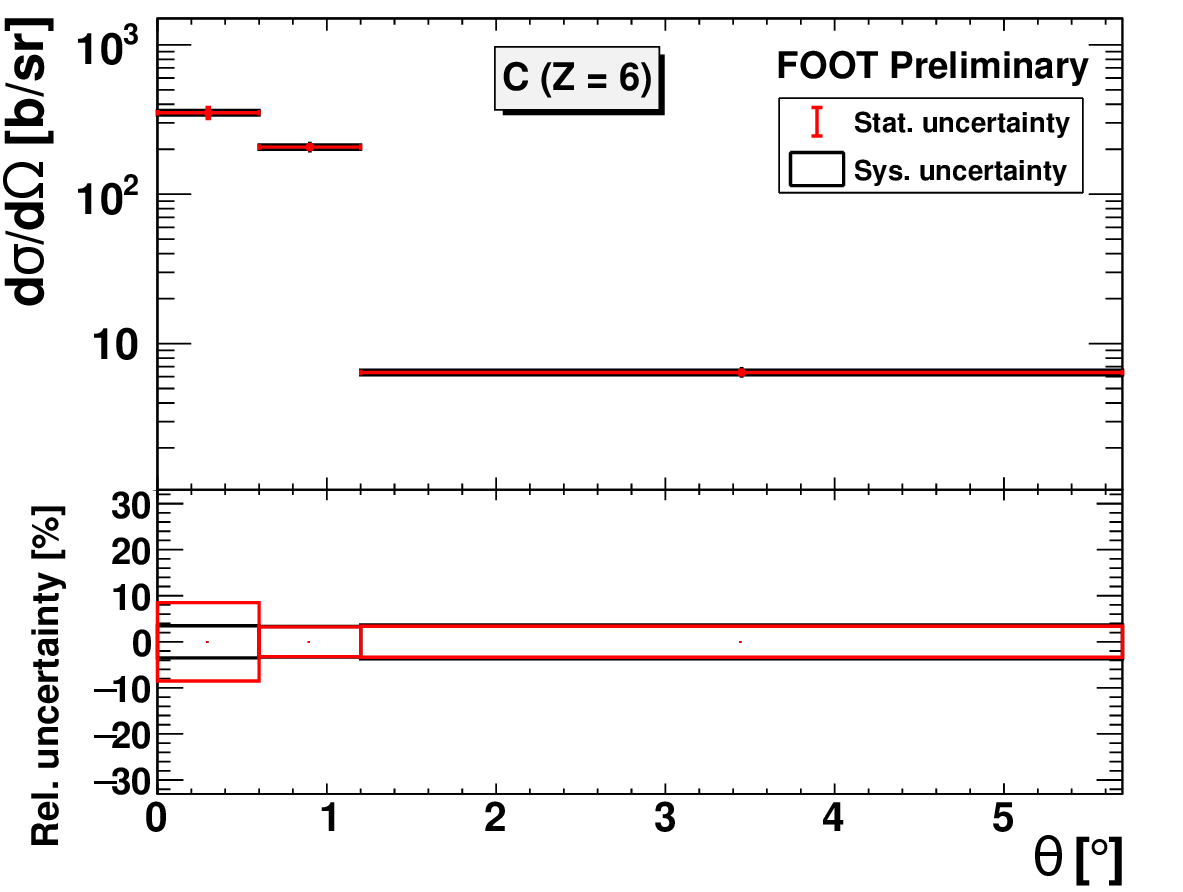}
\end{minipage}
\caption{Angular fragmentation cross sections for a $400\text{MeV/u}$ $^{16}$O beam on a C$_2$H$_4$ target for He (Z=2, left) and C (Z=6, right).}
\label{f.xs_ang}
\end{figure}

\begin{figure}[h]
\centering
\begin{minipage}{0.49\textwidth}
\centering
\includegraphics[width=\textwidth]{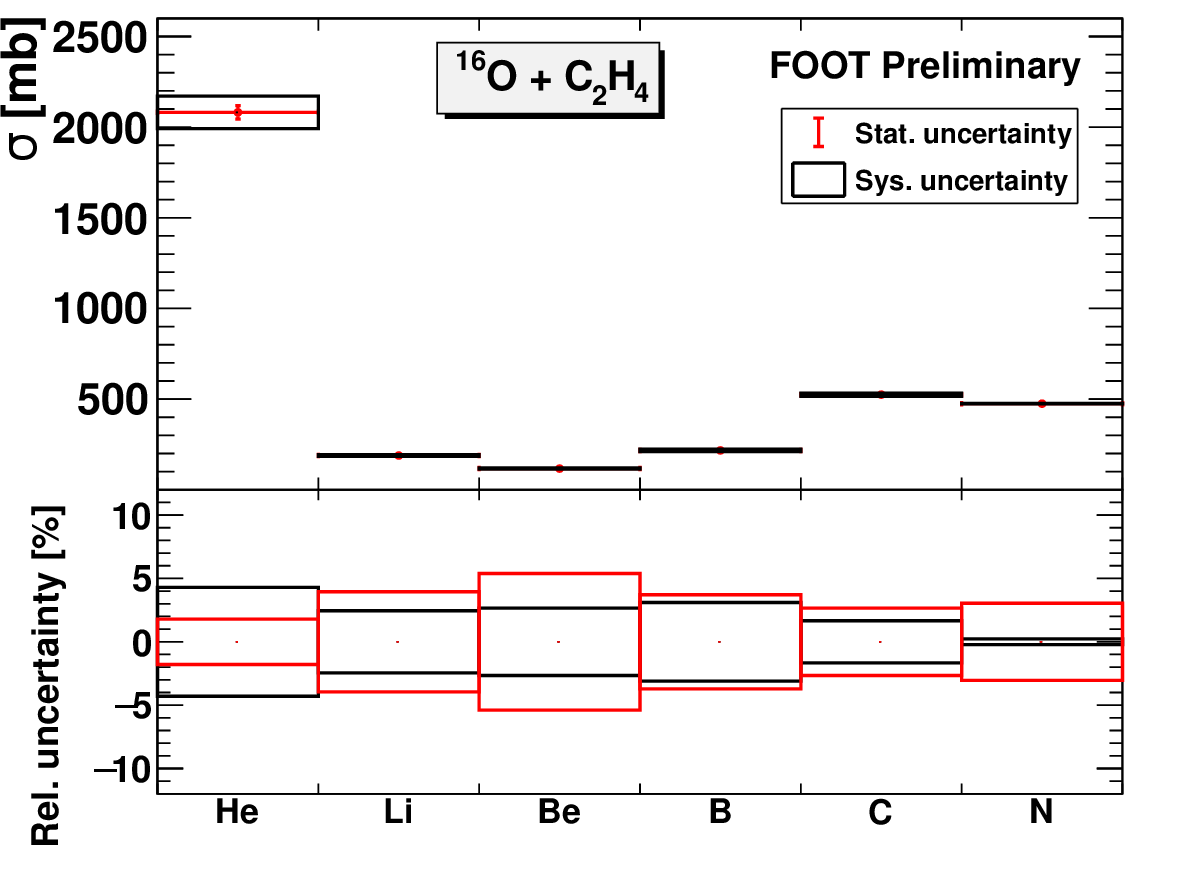}
\end{minipage}
\hfill
\begin{minipage}{0.49\textwidth}
\centering
\vspace{-0.3cm}
\includegraphics[width=\textwidth]{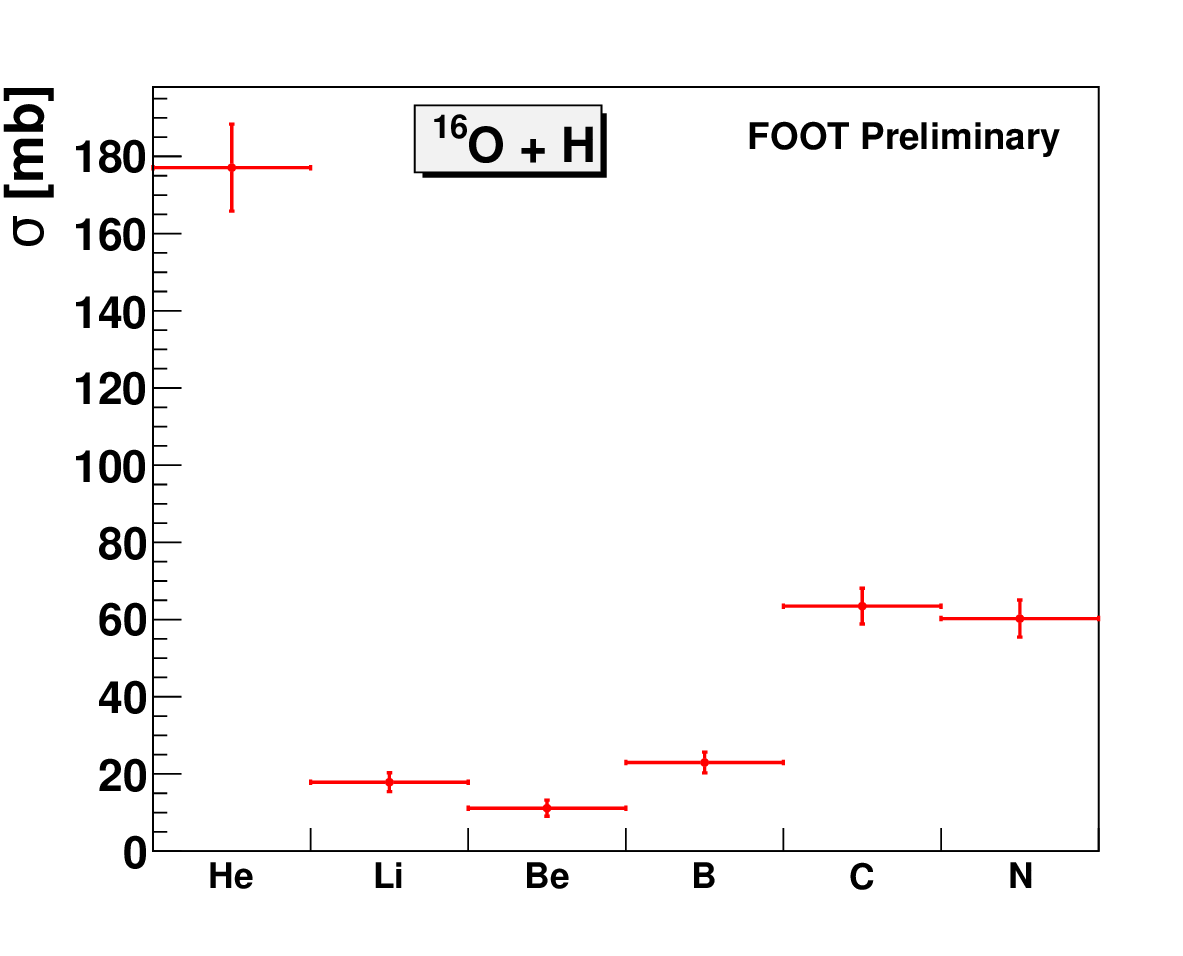}
\end{minipage}
\caption{Total fragmentation cross section for a $400\text{MeV/u}$ $^{16}$O beam on a C$_2$H$_4$ target (left) and a Hydrogen target (right) for different fragment charges.}
\label{f.xs_tot}
\end{figure}

In fig.~\ref{f.xs_ang}, the angular cross sections for a $400\text{MeV/u}$ \(^{16}\)O beam on a C\(_2\)H\(_4\) target for He (Z=2) and C (Z=6) fragments are shown. The systematic uncertainty has been evaluated mainly through MC studies and it accounts for efficiencies and the background subtraction procedure. The relative statistical (in red) and  systematic (in black) uncertainties are shown in the lower part of the plots. In fig.~\ref{f.xs_tot}, the total fragmentation cross section for a $400\text{MeV/u}$ \(^{16}\)O beam on a C\(_2\)H\(_4\) target for different fragment charges is shown on the left, while on the right the total fragmentation cross section for a Hydrogen target, obtained as in eq.~(\ref{e.xs_p}), is presented.
\section{Conclusions}
The analysis presented was performed on data acquired with a reduced part of the electronic setup of the FOOT experiment at GSI in 2021, using a $400\text{MeV/u}$ \(^{16}\)O beam impinging on graphite (C) and polyethylene (C\(_2\)H\(_4\)) targets. Elemental and angular cross sections were obtained for different fragment charges. By subtracting the cross section for a Carbon target from the one for polyethylene, total and angular fragmentation cross sections for a Hydrogen target were also measured. These results demonstrate the feasibility of cross section measurements with the FOOT experiment. In future, the precision of the measurement is expected to notably improve with the complete FOOT setup, allowing also for isotopic cross section measurements.
\\

\newpage
\bibliography{apssamp}

\begin{thebibliography}{5}%
\makeatletter
\providecommand \@ifxundefined [1]{%
 \@ifx{#1\undefined}
}%
\providecommand \@ifnum [1]{%
 \ifnum #1\expandafter \@firstoftwo
 \else \expandafter \@secondoftwo
 \fi
}%
\providecommand \@ifx [1]{%
 \ifx #1\expandafter \@firstoftwo
 \else \expandafter \@secondoftwo
 \fi
}%
\providecommand \natexlab [1]{#1}%
\providecommand \enquote  [1]{``#1''}%
\providecommand \bibnamefont  [1]{#1}%
\providecommand \bibfnamefont [1]{#1}%
\providecommand \citenamefont [1]{#1}%
\providecommand \href@noop [0]{\@secondoftwo}%
\providecommand \href [0]{\begingroup \@sanitize@url \@href}%
\providecommand \@href[1]{\@@startlink{#1}\@@href}%
\providecommand \@@href[1]{\endgroup#1\@@endlink}%
\providecommand \@sanitize@url [0]{\catcode `\\12\catcode `\$12\catcode
  `\&12\catcode `\#12\catcode `\^12\catcode `\_12\catcode `\%12\relax}%
\providecommand \@@startlink[1]{}%
\providecommand \@@endlink[0]{}%
\providecommand \url  [0]{\begingroup\@sanitize@url \@url }%
\providecommand \@url [1]{\endgroup\@href {#1}{\urlprefix }}%
\providecommand \urlprefix  [0]{URL }%
\providecommand \Eprint [0]{\href }%
\providecommand \doibase [0]{https://doi.org/}%
\providecommand \selectlanguage [0]{\@gobble}%
\providecommand \bibinfo  [0]{\@secondoftwo}%
\providecommand \bibfield  [0]{\@secondoftwo}%
\providecommand \translation [1]{[#1]}%
\providecommand \BibitemOpen [0]{}%
\providecommand \bibitemStop [0]{}%
\providecommand \bibitemNoStop [0]{.\EOS\space}%
\providecommand \EOS [0]{\spacefactor3000\relax}%
\providecommand \BibitemShut  [1]{\csname bibitem#1\endcsname}%
\let\auto@bib@innerbib\@empty
\bibitem [{\citenamefont {Schardt}\ \emph {et~al.}(2010)\citenamefont
  {Schardt}, \citenamefont {Elsasser},\ and\ \citenamefont
  {Schulz-Ertner}}]{ref:shulz}%
  \BibitemOpen
  \bibfield  {author} {\bibinfo {author} {\bibfnamefont {D.}~\bibnamefont
  {Schardt}}, \bibinfo {author} {\bibfnamefont {T.}~\bibnamefont {Elsasser}},\
  and\ \bibinfo {author} {\bibfnamefont {D.}~\bibnamefont {Schulz-Ertner}},\
  }\bibfield  {title} {\bibinfo {title} {Heavy-ion tumor therapy: Physical and
  radiobiological benefits},\ }\href
  {https://doi.org/10.1103/RevModPhys.82.383} {\bibfield  {journal} {\bibinfo
  {journal} {Rev. Mod. Phys.}\ }\textbf {\bibinfo {volume} {82}},\ \bibinfo
  {pages} {383} (\bibinfo {year} {2010})}\BibitemShut {NoStop}%
\bibitem [{\citenamefont {Durante}\ and\ \citenamefont
  {Paganetti}(2016)}]{ref:durante}%
  \BibitemOpen
  \bibfield  {author} {\bibinfo {author} {\bibfnamefont {M.}~\bibnamefont
  {Durante}}\ and\ \bibinfo {author} {\bibfnamefont {H.}~\bibnamefont
  {Paganetti}},\ }\bibfield  {title} {\bibinfo {title} {Nuclear physics in
  particle therapy: a review},\ }\href
  {https://doi.org/10.1088/0034-4885/79/9/096702} {\bibfield  {journal}
  {\bibinfo  {journal} {Rep. Prog. Phys.}\ }\textbf {\bibinfo {volume} {79}},\
  \bibinfo {pages} {096702} (\bibinfo {year} {2016})}\BibitemShut {NoStop}%
\bibitem [{\citenamefont {Durante}\ and\ \citenamefont
  {Cucinotta}(2011)}]{ref:cucinotta}%
  \BibitemOpen
  \bibfield  {author} {\bibinfo {author} {\bibfnamefont {M.}~\bibnamefont
  {Durante}}\ and\ \bibinfo {author} {\bibfnamefont {F.~A.}\ \bibnamefont
  {Cucinotta}},\ }\bibfield  {title} {\bibinfo {title} {Physical basis of
  radiation protection in space travel},\ }\href
  {https://doi.org/10.1103/RevModPhys.83.1245} {\bibfield  {journal} {\bibinfo
  {journal} {Rev. Mod. Phys.}\ }\textbf {\bibinfo {volume} {83}},\ \bibinfo
  {pages} {1245} (\bibinfo {year} {2011})}\BibitemShut {NoStop}%
\bibitem [{\citenamefont {Battistoni}\ \emph {et~al.}(2021)\citenamefont
  {Battistoni} \emph {et~al.}}]{ref:battistoni}%
  \BibitemOpen
  \bibfield  {author} {\bibinfo {author} {\bibfnamefont {G.}~\bibnamefont
  {Battistoni}} \emph {et~al.},\ }\bibfield  {title} {\bibinfo {title} {({FOOT}
  collaboration), {Measuring the Impact of Nuclear Interaction in Particle
  Therapy and in Radio Protection in Space: the FOOT Experiment}},\ }\href
  {https://doi.org/10.3389/fphy.2020.568242} {\bibfield  {journal} {\bibinfo
  {journal} {Front. Phys.}\ }\textbf {\bibinfo {volume} {8}},\ \bibinfo {pages}
  {568242} (\bibinfo {year} {2021})}\BibitemShut {NoStop}%
\bibitem [{\citenamefont {Toppi}\ \emph {et~al.}(2022)\citenamefont {Toppi}
  \emph {et~al.}}]{ref:toppi}%
  \BibitemOpen
  \bibfield  {author} {\bibinfo {author} {\bibfnamefont {M.}~\bibnamefont
  {Toppi}} \emph {et~al.},\ }\bibfield  {title} {\bibinfo {title} {({FOOT}
  collaboration), {Elemental} fragmentation cross sections for a
  $^{16}\mathrm{O}$ beam of 400 {MeV/u} kinetic energy interacting with a
  graphite target using the {FOOT $\Delta$E-TOF} detectors},\ }\href
  {https://doi.org/10.3389/fphy.2022.979229} {\bibfield  {journal} {\bibinfo
  {journal} {Front. Phys.}\ }\textbf {\bibinfo {volume} {10}},\ \bibinfo
  {pages} {979229} (\bibinfo {year} {2022})}\BibitemShut {NoStop}%
\end{thebibliography}%

\end{document}